\documentclass[12pt]{article}
\usepackage{cite}
\usepackage{times,color,bm}
\usepackage{epsfig,amsmath,amssymb,amsfonts,mathrsfs,braket}
\usepackage{subfigure,graphicx,epstopdf}
\usepackage{tikz}
\usepackage{amssymb}
\usetikzlibrary{matrix,fit}
\usepackage[colorlinks]{hyperref}
\hypersetup{linkcolor = {blue},	citecolor = {blue},	urlcolor = {blue}}

\makeatletter

\newcommand{\Rmnum}[1]{\expandafter\@slowromancap\romannumeral #1@}
\makeatother

\newenvironment{sciabstract}{%
	\begin{quote} \bf}
	{\end{quote}}

\newcounter{lastnote}

\title{Timestamp Boson Sampling}
\author
{Wen-Hao Zhou,$^{1,2}$ Jun Gao,$^{1,2}$ Zhi-Qiang Jiao,$^{1,2}$ Xiao-Wei Wang,$^{1,2}$\\ Ruo-Jing Ren,$^{1,2}$ Xiao-Ling Pang,$^{1,2}$ Lu-Feng Qiao,$^{1,2}$ Chao-Ni Zhang,$^{1,2}$\\ Tian-Huai Yang,$^{1,2}$ Xian-Min Jin$^{1,2,*}$\\
	\normalsize{$^1$Center for Integrated Quantum Information Technologies (IQIT), School of Physics}\\
	\normalsize{and Astronomy and State Key Laboratory of Advanced Optical Communication Systems}\\
	\normalsize{and Networks, Shanghai Jiao Tong University, Shanghai 200240, China}\\
	\normalsize{$^2$CAS Center for Excellence and Synergetic Innovation Center in Quantum Information and}\\
	\normalsize{Quantum Physics,University of Science and Technology of China, Hefei, Anhui 230026, China}\\
	\normalsize{$^\ast$E-mail: xianmin.jin@sjtu.edu.cn}\\
}
\date{}

\begin{document}
	\baselineskip24pt
	
	\maketitle
	
	\begin{sciabstract}
Quantum advantage, benchmarking the computational power of quantum machines outperforming all classical computers in a specific task, represents a crucial milestone in developing quantum computers and has been driving different physical implementations since the concept was proposed. Boson sampling machine, an analog quantum computer that only requires multiphoton interference and single-photon detection, is considered to be a promising candidate to reach this goal. However, the probabilistic nature of photon sources and inevitable loss in evolution network make the execution time exponentially increasing with the problem size. Here, we propose and experimentally demonstrate a timestamp boson sampling that can reduce the execution time by 2 orders of magnitude for any problem size. We theoretically show that the registration time of sampling events can be retrieved to reconstruct the probability distribution at an extremely low-flux rate. By developing a time-of-flight storage technique with a precision up to picosecond level, we are able to detect and record the complete time information of 30 individual modes out of a large-scale 3D photonic chip. We successfully validate boson sampling with only one registered event. We show that it is promptly applicable to fill the remained gap of realizing quantum advantage by timestamp boson sampling. The approach associated with newly exploited resource from time information can boost all the count-rate-limited experiments, suggesting an emerging field of timestamp quantum optics.
		\\
	\end{sciabstract}	

Quantum computing, based on the superposition, entanglement, and interference principles of quantum states, has the potential capability to outperform any existed classical computers for certain problems, attracting extensive attention and research to develop quantum algorithms~\cite{qs1,qs2} and physically construct quantum computers~\cite{qc}. However, the noise in quantum gates and the decoherence in quantum systems limit the size of the quantum circuits and the number of the controllable qubits, which make the realization of universal quantum computers still be challenging and considered as a long-term goal~\cite{challenge}. As an intermediate but crucial milestone for benchmarking the development of quantum computing, quantum advantage~\cite{sm,sm2}, means that a quantum machine can perform a specific task demonstrating computational power beyond the state-of-art classical computers. Many efforts have been devoted to construct quantum systems \cite{qubit1,qubit2,qubit3,qubit4} with a state space large enough to meet the required size of the quantum machine.

Boson sampling, first proposed by Aaronson and Arkhipov~\cite{bs}, is to sample the output probability distribution of $n$ identical bosons scattering in an $m$-mode ($m>n$) interferometer, which can be expressed by a Haar random matrix $\emph{U}$. The computational complexity of boson sampling is quantified by the calculation of the submatrix permanent of \emph{U}, which can be further enhanced by the combination factor $\dbinom{m}{n}$~\cite{enhance1,enhance2}. From the experimental point of view, in the photonic scheme, only single photon sources, linear-optical network and single-photon detection are needed to realize boson sampling. Due to the theoretically well-defined and experimentally friendly feature, boson sampling is considered to be a short cut to achieve quantum advantage. 

Up to now, boson sampling has been realized in many scenarios ~\cite{qubit3,bse1,bse2,bse3,bse4,bse5,bse6,bse7,bse8,bse9,bse10}, and variant protocols such as scattershot boson sampling~\cite{scattershot1,scattershot2} and Gaussian boson sampling~\cite{guassion} have also been conducted. Simulations on supercomputer have been shown that 50 photons or equivalent size of quantum systems may beat the current calculation limit, so as to achieve quantum advantage~\cite{classical,classical2}. However, we are far away from accessing such a system scale, since the inefficiency in collecting correlation events leads to exponentially increased execution time due to the probabilistic nature of single photon sources and inevitable loss in linear-optical network. 

Statistical errors caused by Poisson distribution are magnified in the extremely low-flux rate, which is not only an essential restriction for boson sampling but also for all the count-rate-limited experiments. In general, we require hundreds of counts to reduce this statistical error, but in reality, to achieve quantum advantage, output combinations $\dbinom{m}{n}$ of boson sampler must be very huge, leading to a very low-flux rate for every combination. Thus, most combinations may only be detected a few times, even only once, which prevents a standard boson sampling from realizing in such a situation.

In order to solve the problems above, we propose a timestamp boson sampling which can even be operated in the extremely low-flux rate and reduce the experimental time consumption by 2 orders of magnitude in any problem size. We retrieve the registration time $\tau_i$ of each combination as the timestamp and reconstruct the probability based on the principle that the earlier the arrival, the higher the probability. The reconstruction probability $p_i$ of each combination can be described by Eq.1 and the output distribution satisfies the normalization $\sum_{i}{p_i}=1$.
\begin{equation}
   p_i=\frac{\tau_i^{-1}}{\sum_{i}\tau_i^{-1}}
	\label{eq:1}
	\end{equation} 

For recording the complete time information of all the individual modes, we develop a time-of-flight storage (ToFS) technique, which has been shown in Fig.1. The ToFS is mainly composed of two parts: the trigger channel and the detection channels. The delay of 30 detection channels can be set independently and the time information of all 31 channels is recorded simultaneously (see Supplementary Materials Fig.S1). The time accuracy of individual channel reaches the picosecond level and such a high time resolution guarantees the precision of the reconstructed probability distribution. We extract the fourfold coincidence events within the coincidence time window (see Methods) and use the time information $\tau_i$ of the trigger as the timestamp to mark the occurrence of each combination.
	
The linear-optical network is realized by 3D photonic chip fabricated by femtosecond laser direct writing technique~\cite{fs1,fs2,qw,fast,qt} (see Methods for fabrication details). The waveguide structure inside the chip is shown in Fig.2(a), where 6 input waveguides are coupled into a 2D waveguide array and then the 3D $5 \times 6$ structure transforms into a tiled 30 output waveguides array after evolution. The spacing of all the waveguides for input and output is set to 127$\mu m$ in order to match the external V-groove structure. To introduce random phase shift~\cite{bse6}, we deform the S-bend randomly of each input and output waveguide. 
		
We use single photon scattering and a series of HOM interferences to reconstruct the unitary matrix $\emph{U}$. The measured results of the amplitude elements and the phase elements are shown in Fig.2(b) and 2(c) (see Methods for characterization details). In Fig.2(b), to obtain all the moduli of the matrix, we inject heralded single photons into all 6 inputs one by one and measure all 30 output modes simultaneously. In Fig.2(c), the phase relationship between different output modes is achieved by measuring hundreds of HOM interferences using the selected input ports~\cite{ch} (See Supplementary Materials Fig.S2).
	
Besides the fabrication and full characterization of the 3D photonic chip, we build a 6-photon source which is depicted in details in Fig.3(a). A mode-locked Ti:sapphire laser emits a femtosecond pulse (140fs) with the center wavelength of 780nm and then is frequency-doubled by a lithium triborate (LBO) crystal. The generated 390nm laser is split into two spatial modes by a balanced UV beam splitter. The reflected laser passes through two beta barium borate (BBO) crystals and generates two pairs of correlated photons via spontaneous parametric down conversion. The BBO1 crystal is type-$\rm\Rmnum {2}$ phase matching in a beam-like scheme~\cite{beamlike1,beamlike2} while the BBO2 crystal is type-$\rm\Rmnum{2}$ collinear phase-matched configuration~\cite{colline}. The transmission laser passes through the BBO3 crystal which is the same as the BBO1. At last, photons are coupled into PM optical fibers connected to the PM V-groove fiber array. We achieve the temporal overlap of non-classical interference among different down-converted photons by translating linear motorized stages to compensate external delays.
	
We report the experimental results in two different input scenarios. One is a 4-photon  experiment with 3-photon injection and 1 external trigger, whose sampling rate is sufficient so that we can detect almost all the output combinations within a moderate accumulation time. The other is a 6-photon experiment with 3 heralded single photons injection, and we can only detect parts of the output combinations due to the low-flux rate. It is worth mentioning that in the 6-photon case, we synchronize the coincidence of 3 trigger signals into 1 signal with a field programmable gate array (FPGA) module, so that we regard the synthesized triggers as the timestamps, but in essence, they are sixfold coincidence events. 

For the 4-photon experiment, we collect a total of 50000s data when the pump power is set at 600mW. We observe 83455 fourfold coincidence events in total with a smpling rate about 6000 per hour. We divide the total time into 6000 intervals and count the event amounts in each time interval. The result has been shown in Fig.3(b), which approximates a Poisson distribution. The probabilistic of registration events lead to statistical errors in standard boson sampling protocol, especially when the sampling rate is extremely low.

We normalize the probability distribution using both counting and timestamp protocol. To reduce the data singular points caused by the fluctuation of registration time, we use the average timestamp of each combination occurring 5 times to reconstruct the probability distribution (also see more experimental results with different occurrence number in Supplementary Materials Fig.S3). We plot the timestamp reconstruction results ($p_i$) and counting reconstruction results ($c_i$) in a five-layer ring, as shown in Fig.3(c) (see Methods for processing details). The similarity between the two reconstruction protocols $S=\sum_{i}\sqrt{p_ic_i}$ is up to 98.7\%. The total variation distance is defined as $D=(1/2)\sum_{i}\left|p_i-c_i\right|$ and the value is 0.13. These two reconstruction results are highly similar. In stead of consuming 50000s data and 83455 events to realize the counting reconstruction, we only use an average of 5 events to achieve the same goal in timestamp boson sampling. The execution time and the required sampling rate are reduced by two orders of magnitude, indicating a great resource-effective advantage of our protocol.

When it is infeasible to detect all the combinations in large-scale standard boson sampling, we should still be able to show a valid validation~\cite{v1,v2,v3} even with a small number of samples. For 6-photon experiment, after a collection time of nearly 100 hours, we obtain 358 sixfold coincidence events in total, where 321 combinations just happen once, 17 combinations happen twice and only 1 combination happens three times. The counting reconstruction suffers huge statistical error, which cannot be considered as the data supporting an effective validation. 

Instead, we use the timestamp boson sampling protocol and validate the reconstruction results~\cite{v1,v2,v3,v4,v5} with only one registered event. To distinguish our experimental results from the uniform distribution, we perform the row-norm estimator test~\cite{v2}, and the results are shown in Fig.3(d). The different growth trend between the experimental data and the simulated uniform data indicates that the reconstruction results are not extracted from a uniform distribution. We also focus on verifying whether our experiment results are based on indistinguishable photon scattering by performing the likelihood ratio test~\cite{v3}. The experimental data and the simulated distinguishable data show a distinctly different growth trend in Fig.3(e), which ensures that our experimental data are sampled by indistinguishable photons. Timestamp reconstruction protocol can still work properly even in the extremely low-flux and low-sampling rate, and can be very useful for large-scale validation against classical imposters.

We further mine more information about the nature of the timestamp reconstruction. In Fig.4(a), we extract the time interval of both 83455 events and fix output configuration (1,10,22). We can find that the time interval between successive occurrences of events satisfying exponential decay distributions. Compared with Figure.3(b), although their distributions are different, they share the same nature because the physical description of boson sampling processes are equivalent. Timestamp protocol provides a new way for us to analyze the same physical process.

For finding a suitable collection region and getting rid of the singular points simultaneously, we statistically analyze the relationship between the number of occurrences and the number of singular points, see the results of 4-photon experiment as an example in Fig.4(b). When each combination only occurs once, a large number of singular points will be discarded. Until the number of occurrences is increased from 5 to 10, the proportion of the discarded singular points drops from 9.96\% to 1.77\%. We can obtain a good tradeoff between the success rate and the execution time in timestamp boson sampling experiment.

Fig.4 (c)-(h) show the comparison between the experimental ($p_i$) and the theoretical ($t_i$) results with the timestamp protocol (The corresponding validation results can be found in Supplementary Materials Fig.S4). We use fidelity and total variation distance to quantify the performance of our experimental result, which are defined by $F=\sum_i\sqrt{p_it_i}$ and $D=(1/2)\sum_{i}\left|p_i-t_i\right|$. For a perfect boson sampler, $\emph{F}$ should be 1 and $\emph{D}$ should be 0. By adding the number of occurrences one by one, we can observe the fidelities and total variation distance slightly improve, but deviate from the theoretical predictions to some extent. The deviation is attributed to the additional loss of the tiled 30 output waveguides array, which can be eliminated by developing imaging-based large-scale correlation measurement~\cite{camera} in the future. 
	
With the newly added resource, it is worth estimating how the timestamp protocol help to approach quantum advantage. The computational complexity of the boson sampling problem is related to the calculation of the submatrix permanent of $\emph{U}$. As a \#P-hard problem, even with the best known classic algorithm, the computational steps still need $\emph{O}(n2^n)$~\cite{O}, where $n$ represents the number of photons. In fact, there has been no conclusion about the minimum $n$ to achieve quantum advantage, from $n=7$~\cite{enhance1}, $n$ between 20 and 30~\cite{bs}, to $n=50$~\cite{50}, and here we choose the maximum benchmark $n=50$ for our estimation. If we consider the contribution of all combinations $\dbinom{m}{n}$ to the computational complexity, it will take $\emph{O}(n2^n)\dbinom{m}{n}$ equivalent computational steps, where $m$ represents the number of modes. In general, it is commonly believed that $m$ needs to be much larger than $n^2$ in order to meet the permanent-of-Gaussians conjecture (at least $m=n^2$). However, another view is that this condition can be extended to $m=\emph{O}(n)$, such as $m=2n$ and it can also achieve quantum advantage, and we choose a quite stringent condition $m=2n$ for our estimation. 

In Fig.5, we calculate the lower bound of the quantum advantage with the formula $\emph{O}(n2^n)$ and the quantum advantage region is shown in gray area. The excitation rate of the best known quantum dot source is about 0.6~\cite{cc} and the number of photons with the largest system scale is about 15~\cite{qubit3}. Based on these parameters, we present the required total efficiency (see Methods for calculation details) under different photon number and mode number. We then show the enhancement of photon number and mode number further enabled by our timestamp protocol. Meanwhile, we give all the computational steps with the formula $\emph{O}(n2^n)\dbinom{m}{n}$. To reach quantum advantage, the standard boson sampling protocol require a total efficiency as high as 0.9, but timestamp boson sampling only require a moderate total efficiency of 0.68 instead. We can see that timestamp boson sampling prompt accessing the computational complexity by 2 to 3 orders of magnitude and therefor fills up the key gap between current capability and quantum advantage.  

In summary, we propose and experimentally demonstrate a timestamp boson sampling based on the retrieval of the registration time information of sampling events. Experimental results show that the timestamp reconstruction distribution is highly similar to the counting reconstruction up to 98\%, but reduce the execution time by 2 orders of magnitude. The validation results verify that the experimental events are still from a genuine boson sampler even with only one registered event. Further analyzation also shows that timestamp boson sampling can be used to increase the number of both photons and modes additionally from the state of art, thus is able to fill the remained gap of realizing quantum advantage. Moreover, we show that it is possible to obtain faithful results from very limited events by retrieving their time information. Such a mechanism can be generalized to a wide range of quantum optics, such as multiphoton quantum walk~\cite{mqw}, noisy intermediate-scale quantum (NISQ) technology~\cite{challenge}, and multiphoton entanglement~\cite{qubit2}, driving our capacity of controlling quantum systems to an entirely new accessible realm.\\
	
	\subsection*{Acknowledgments}
The authors thank Jian-Wei Pan for helpful discussions. This work was supported by National Key R\&D Program of China (2019YFA0308700 and 2017YFA0303700); National Natural Science Foundation of China (NSFC) (61734005, 11761141014, 11690033); Science and Technology Commission of Shanghai Municipality (STCSM) (17JC1400403); Shanghai Municipal Education Commission (SMEC) (2017-01-07-00-02-E00049); X.-M. J. acknowledges additional support from a Shanghai talent program.

\subsection*{Data availability.}
The data that support the findings of this study are available from the corresponding author upon reasonable request.
\\

	\subsection*{Methods}

	\paragraph*{Time-of-flight storage technique}
The main part of the ToFS has been shown in Fig.1. The output part of the photonic chip is connected to 30 avalanche photodetectors (APD) with a multimode V-groove fiber array. The detector signals are linked to the 30 channels time-of-flight storage module. We add the trigger signal and set the delay of each channel relative to the trigger signal respectively. We scan the delay between different detection channels and the trigger, which identify the differential delays by maximizing their coincidence counts. The coincidence time window is set as 2ns, and then we get fourfold coincidence events through post-processing. In order to mark each fourfold coincidence, we use the time information $\tau_{i}$ of each trigger signal as the timestamp and $\tau_i$ is related to the reciprocal of counting information ${N_i}^{-1}$.
	
	\paragraph*{Fabrication of 3D photonic chip}
The 3D photonic chip is fabricated by a femtosecond laser with the repetition frequency of 1MHz, the pulse duration of 290fs and the wavelength of 513nm. After being reshaped by a cylindrical lens, the writing laser is focused into a borosilicate glass substrate with a $\times$50 objective lens (NA=0.55). The sample is placed on a high-precision three-axis translation stage moving at a velocity of 15mm/s. We obtain a photonic chip with 1D 6-input fan-in, 30-output fan-out and 3D $5\times6$ random coupling structure. All the inputs and outputs are set at 170 $\mu m$ underneath the surface of the chip substrate. In addition, we introduce randomness by changing different coupling length and introducing S-bend. The input waveguides are coupled to a polarization-maintaining (PM) V-groove fiber array and the output waveguides are coupled to a multimode V-groove fiber array.

	\paragraph*{Characterization of 3D photonic chip}
The unitary scattering process of the 3D photonic chip can be described by the reconstruction of a unitary matrix $\emph{U}$, and each element in the matrix is a complex number. In order to determine the real part of the elements, we inject heralded single photons and measure the intensity of each output mode, and then we can easily obtain these values. Considering the chip efficiency and randomness, we choose port 2, 4 and 5 as the input ports. The determination of the imaginary part can be divided by two steps: firstly, we determine the absolute value by measuring the visibility of two-photon HOM interference, secondly, we determine the sign of imaginary part by scanning and comparing a large number of HOM interferences between different channels. Finally, a $3 \times 30$ matrix can be reconstructed by using the characterization data obtained above.
	
	\paragraph*{Timestamp reconstruction protocol} 
Taking the 4-photon experiment as an example, we illustrate our reconstruction steps when the number of occurrences equals to 1. Statistics show that 4021 out of 4060 combinations happen at least once.

Step $\rm\Rmnum{1}$. We reconstruct the probability distribution of all 4021 combinations by using their time information $\tau$ denoted by the detection time of the trigger.

Step $\rm\Rmnum{2}$. We calculate the theoretical occurrence number $N_\tau$, which is defined as $N_\tau=T/\tau$. $T$ represents the total time 50000s and $\tau$ represents the timestamp of each combination. We also count the actual occurrence number $N_a$ of each combination in total time.

Step $\rm\Rmnum{3}$. According to Poisson distribution, $N_a$ fluctuates but highly stays in the range of $\big[N_a-\sqrt{N_a} , N_a+\sqrt{N_a}\big]$. For low-flux rate scenario, we use more tight but appropriate bound to reshape our data and only keep the events in the range $N_a/2<N_\tau<2N_a$. We normalize the data in the subspace and compare it with the distribution obtained from theory and standard boson sampling.

We process the 4-photon and 6-photon experimental data with the same procedure. The only difference is that when the number of occurrences increases, we replace the single timestamp with the average timestamp of multiple events, which makes the data processing more precise.

	\paragraph*{Computational steps calculation}
The sampling rate of n-photon boson sampling can be expressed by
	\begin{equation}
	SR=\frac{R_{pump}}{n}\eta^n{\dbinom{m}{n}}^{-1}
	\label{eq:2}
	\end{equation}
where $R_{pump}$ represents the pumping repetition rate of the single-photon source, $\eta$ represents the total efficiency of each channel in the device, including the single-photon excitation rate, fiber coupling efficiency, detection efficiency, etc., $n$ represents the number of photons and $m$ represents the number of modes. Considering the best state-of-art parameters of single-photon excitation rate~\cite{cc} and $n$~\cite{qubit3}, and supposing all other efficiencies equal to 1, we set the benchmark at $\eta=0.6$ and $n=15$. In the case of the same sampling rate, we calculate the total efficiency $\eta$ under different $n$ and $m$ by using Eq.(\ref{eq:2}), where $m=2n$. Then we get the computational steps $\emph{o}(n2^n)\dbinom{m}{n}$. In our protocol, we can reduce the sampling rate by 2 orders of magnitude, so according to the equation $SR(n)=100SR(n')$ under the same $\eta$, we estimate the equivalent photon number $n'$ and the corresponding computational steps. Moreover, we also calculate the bound of the quantum advantage by using the formula $\emph{O}(n2^n)$ and $n$ is chosen as 50.
	
	\clearpage
	
\clearpage
	
	\begin{figure}[htbp]
		\centering
		\includegraphics[width=1.0\linewidth]{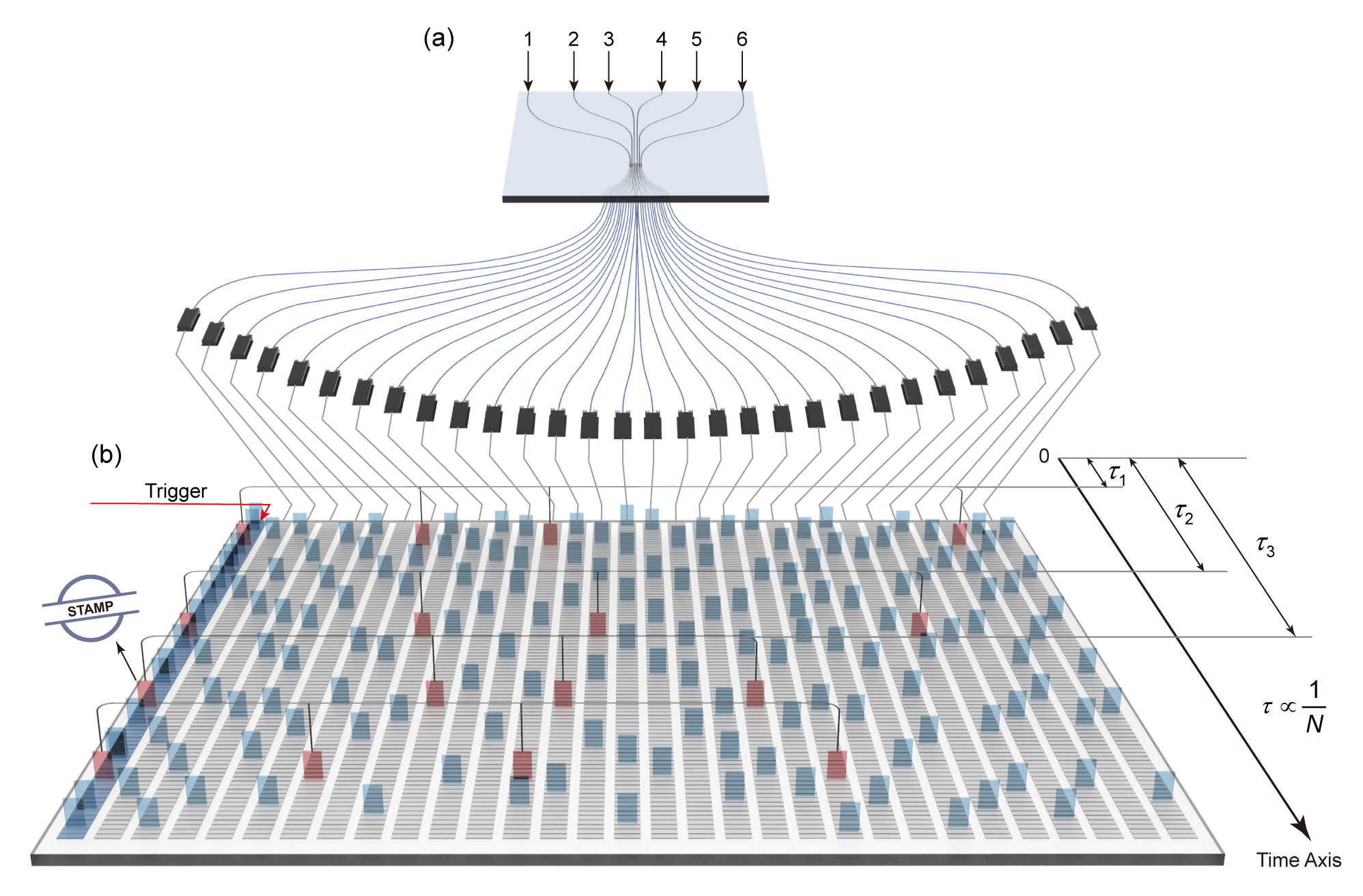}
		\caption{\textbf{Sketch of timestamp boson sampling protocol.} \textbf{(a)} The output ports of the 3D photonic chip are connected with the detector arrays through a multimode V-groove fiber array, then the detected signals are linked into the 30 channels time-of-flight storage (ToFS) module. \textbf{(b)} The ToFS module is mainly composed of the trigger channel (blue background) and the detection channels (grey background). Each channel can set delay freely to compensate the optical path differences between different modes and record all the time information (blue cards). Fourfold coincidence events (red cards) are obtained through post-processing within a 2ns coincidence window. We use the trigger registered time $\tau_{i}$ as the timestamp to mark each coincidence event.}
		\label{fig.1}
	\end{figure}
	
	\clearpage
	
	\begin{figure}[htbp]
		\centering
		\includegraphics[width=0.9\linewidth]{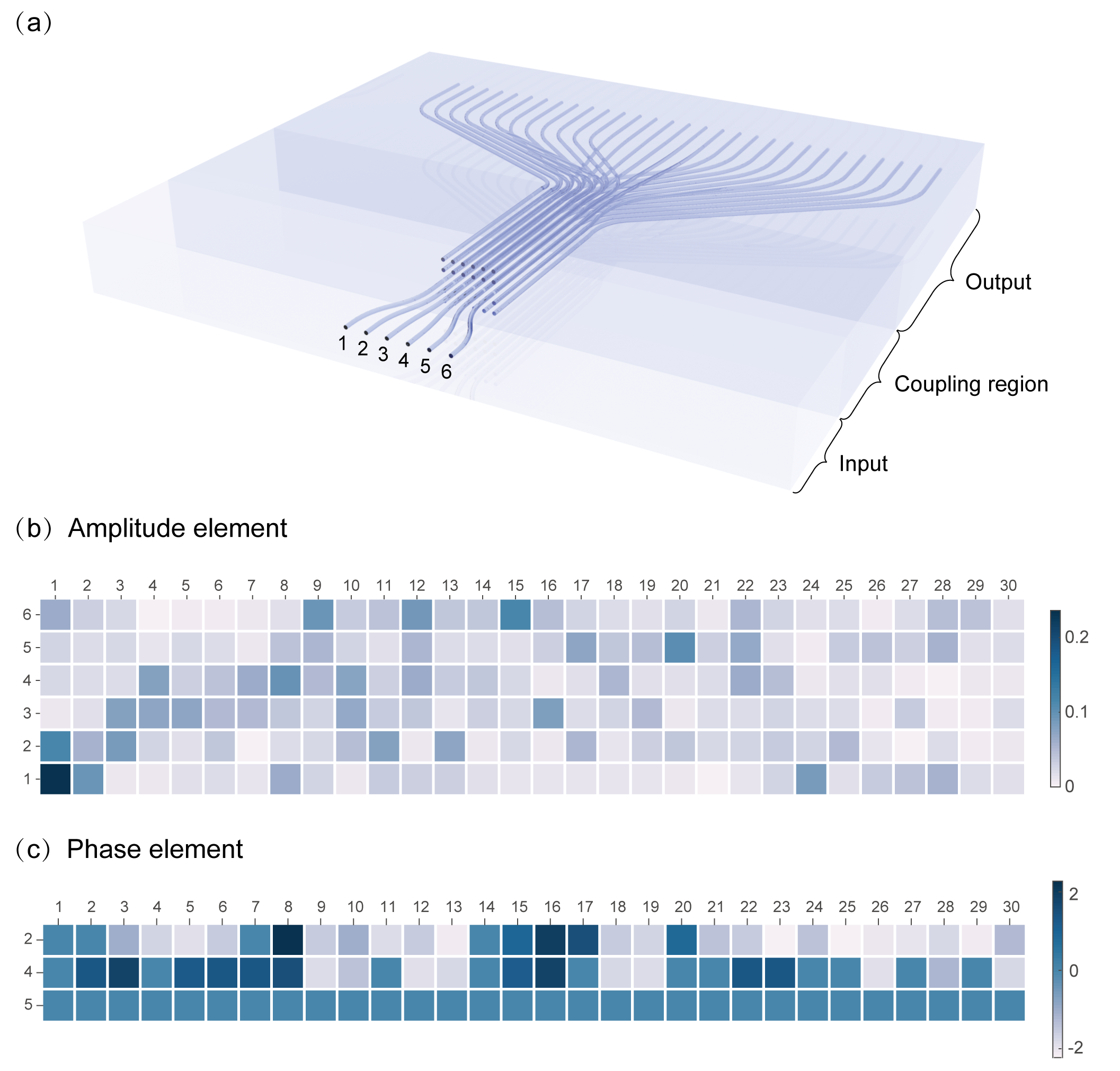}
		\caption{\textbf{Chip structure and scattering matrix characterization.} \textbf{(a)} The 3D photonic chip consists of 6 inputs, 30 outputs and 3D 5$\times$6 random coupling structure. \textbf{(b)} The amplitudes of the scattering matrix with 6 inputs are characterized by injecting heralded single photons into each input. Port 2, 4 and 5 are chosen as the injection ports for boson sampling experiments. \textbf{(c)} The phase elements are measured by hundreds of HOM interferences using the selected input ports.}
		\label{fig.2}
	\end{figure}
	
	\clearpage
	
	\begin{figure}[htbp]
		\centering
		\includegraphics[width=1.0\linewidth]{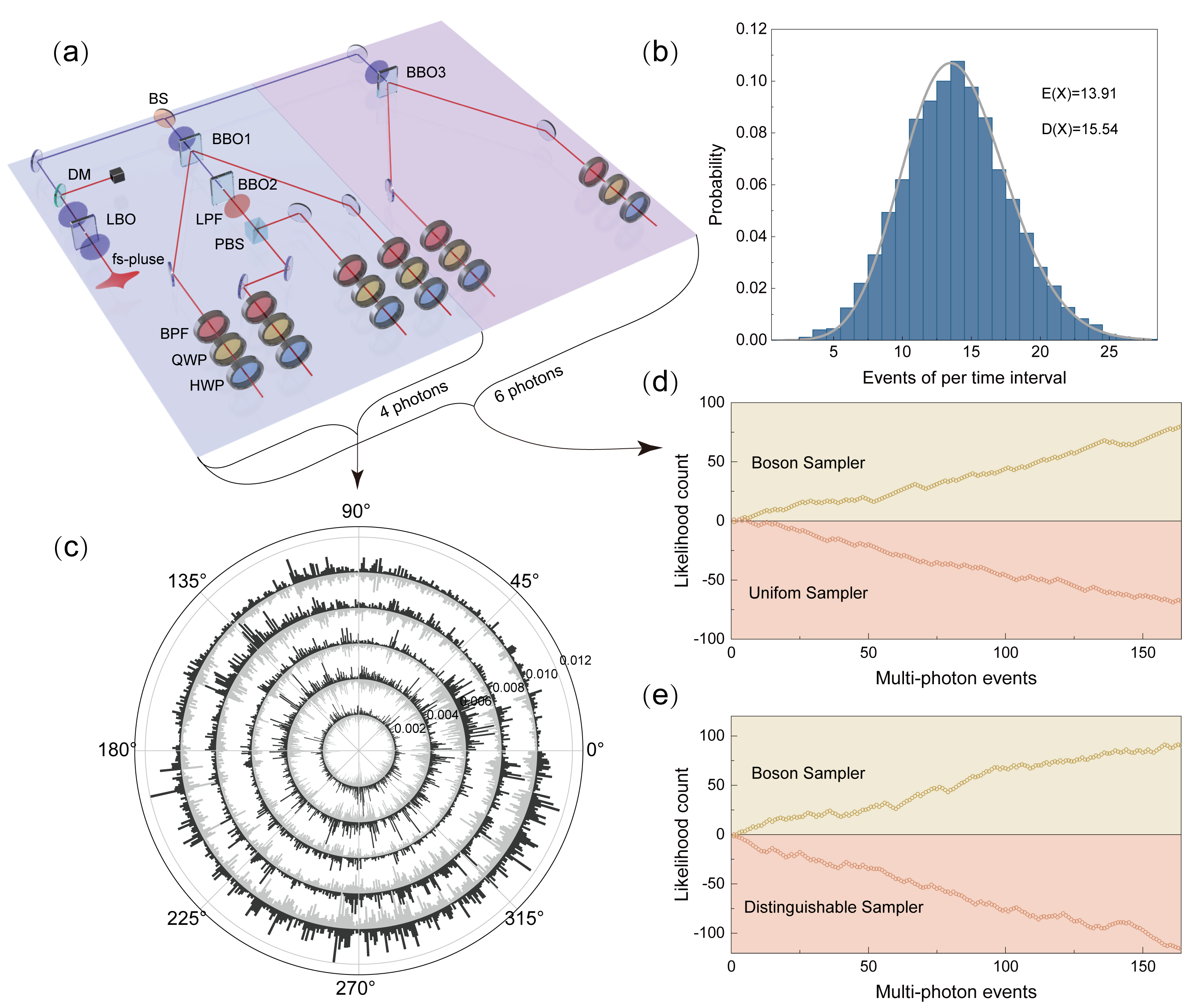}
		\caption{\textbf{Experimental timestamp boson sampling.} \textbf{(a)} Setup of 6-photon source. BBO1 and BBO3 are type-$\rm\Rmnum{2}$ phase matching in a beam-like scheme. BBO2 is type-$\rm\Rmnum{2}$ collinear phase-matched configuration. The 3nm bandpass filters (BPF) are employed to guarantee the spectrum of each path approximately the same. The half waveplate (HWP) and quarter waveplate (QWP) are used to initialize the polarization. \textbf{(b)} The distribution of occurrence times. The statistical approximation is Poisson distribution, whose mathematical expectation $E(x)=\sum_{j}e_{j}\rho_{j}$ equals to 13.91 and the variance $D(x)=\sum_{j}(e_{j}-\bar{e})^2\rho_{j}$ equals to 15.54, where $e_{j}$ represents the number of events in per time interval, corresponding to the probability of $\rho_{j}$, and $\bar{e}$ is the average of $e_{j}$. \textbf{(c)} 4-photon probability distribution comparison between the timestamp reconstruction $p_i$ (dark grey) and the counting reconstruction $c_i$ (light grey). The similarity $S=\sum_{i}\sqrt{p_ic_i}$ is up to 98.7\%. The total variation distance $D=(1/2)\sum_{i}\left|p_i-c_i\right|$ is 0.13. \textbf{(d-e)} The validation of 6-photon timestamp boson sampling. The 6-photon experimental data are performed by both the row-norm estimator test and the likelihood ratio test with only one registered event, both well deviating from uniform and distinguishable sampler. The yellow dots represent experiment data and the red dots represent the simulated data.}
		\label{fig.3}
	\end{figure}

	\clearpage
	
	\begin{figure}[htbp]
		\centering
		\includegraphics[width=1.0\linewidth]{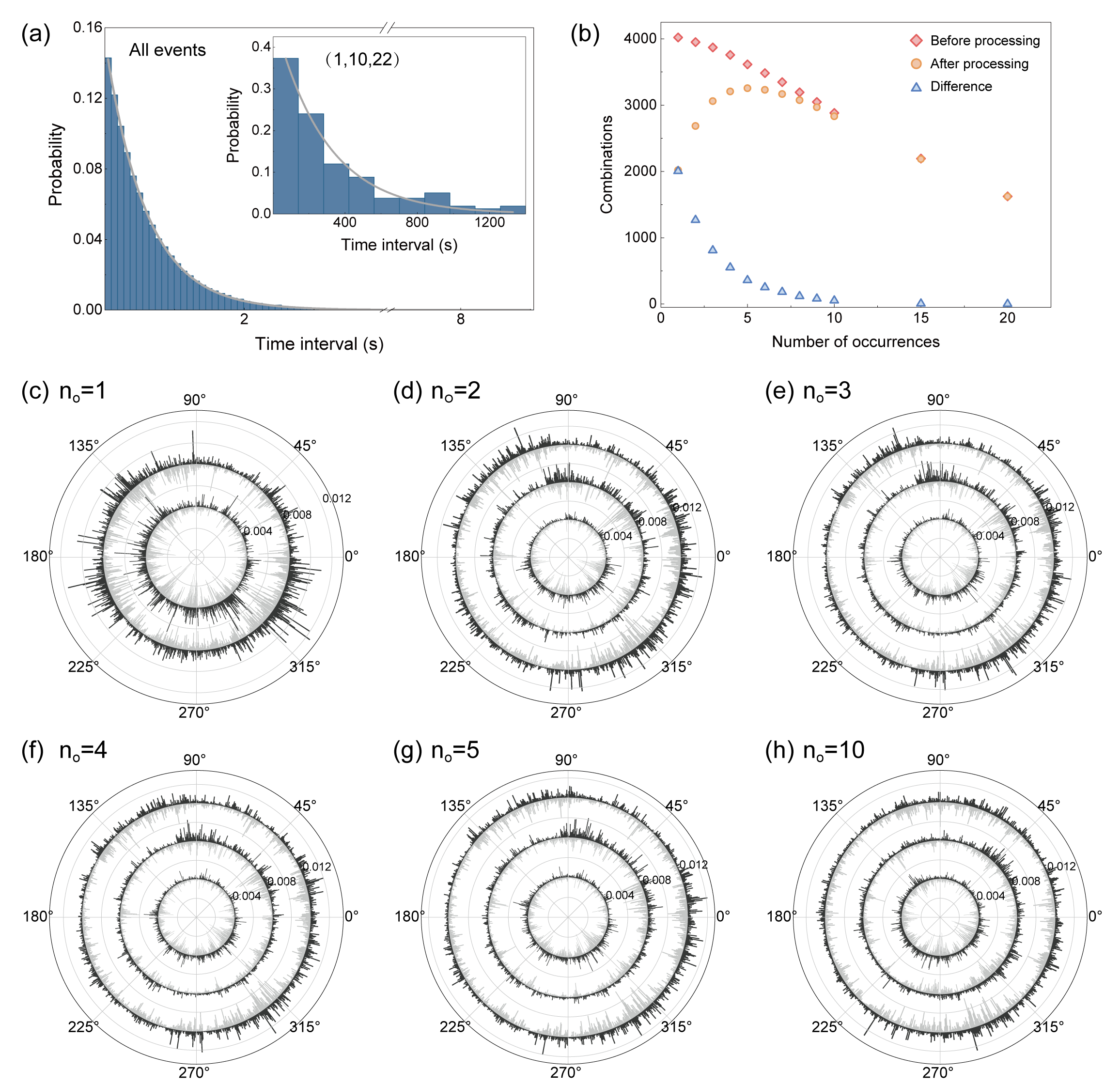}
		\caption{\textbf{Optimization by adding events one by one.} \textbf{(a)} The time interval between successive occurrences of events meets exponential decay distributions. \textbf{(b)} Statistics on the number of occurrences of different combinations before and after data processing. The difference represents the singular points of the data. Taking the average timestamp that each combination occurs  5 to 10 times can effectively eliminate the singular points and improve the accuracy of data processing. \textbf{(c-h)} Experimental $p_i$ (dark grey) and theoretical $t_i$ (light grey) distributions of different occurrence times $n_o$ with the timestamp reconstruction. The calculated fidelity $F=\sum_i\sqrt{p_it_i}$ are 87.7\%, 88.3\%, 88.4\%, 88.6\%, 88.5\% and 89.1\%. The corresponding total variation distance $D=(1/2)\sum_{i}\left|p_i-t_i\right|$ are 0.383, 0.370, 0.368, 0.367, 0.371 and 0.360.}
		\label{fig.4}
	\end{figure}
	
	\clearpage

	\begin{figure}[htbp]
		\centering
		\includegraphics[width=1.0\linewidth]{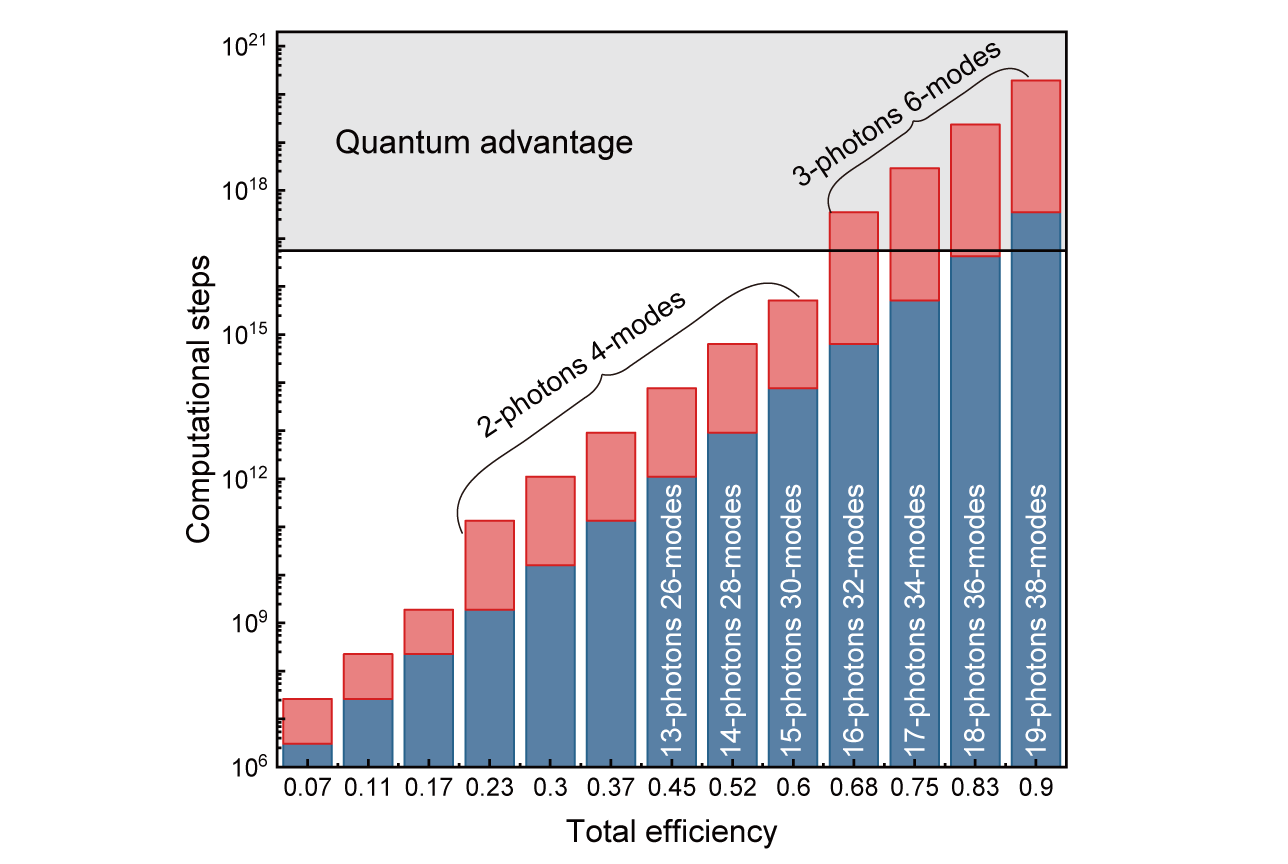}
		\caption{\textbf{Contribution towards quantum advantage.} The region of quantum advantage is shown in gray area, where the lower bound is calculated with the formula $\emph{O}(n2^n)$ and the photon number $n$ is chosen as 50. In light of the best state-of-art total efficiency $\eta=0.6$, we show the required total efficiency for different photon number and mode number in the blue parts. By using the blue parts as benchmarks, the red parts represent the enhancement of $n$ and $m$ with the timestamp protocol. All the computational steps are calculated by $\emph{O}(n2^n)\dbinom{m}{n}$. When the total efficiency touches 0.68, timestamp protocol will be able to fill up the key gap of 3 orders of magnitude in computational complexity between practical experiment and quantum advantage. In stead, standard boson sampling will not be able to achieve quantum advantage if total efficiency goes below $\eta=0.9$.}
		\label{fig.5}
	\end{figure}
	
	\clearpage
	
\section{Supplementary Materials: More preparation work and experimental results}
\begin{figure}[htbp]
		\centering
		\includegraphics[width=1\linewidth]{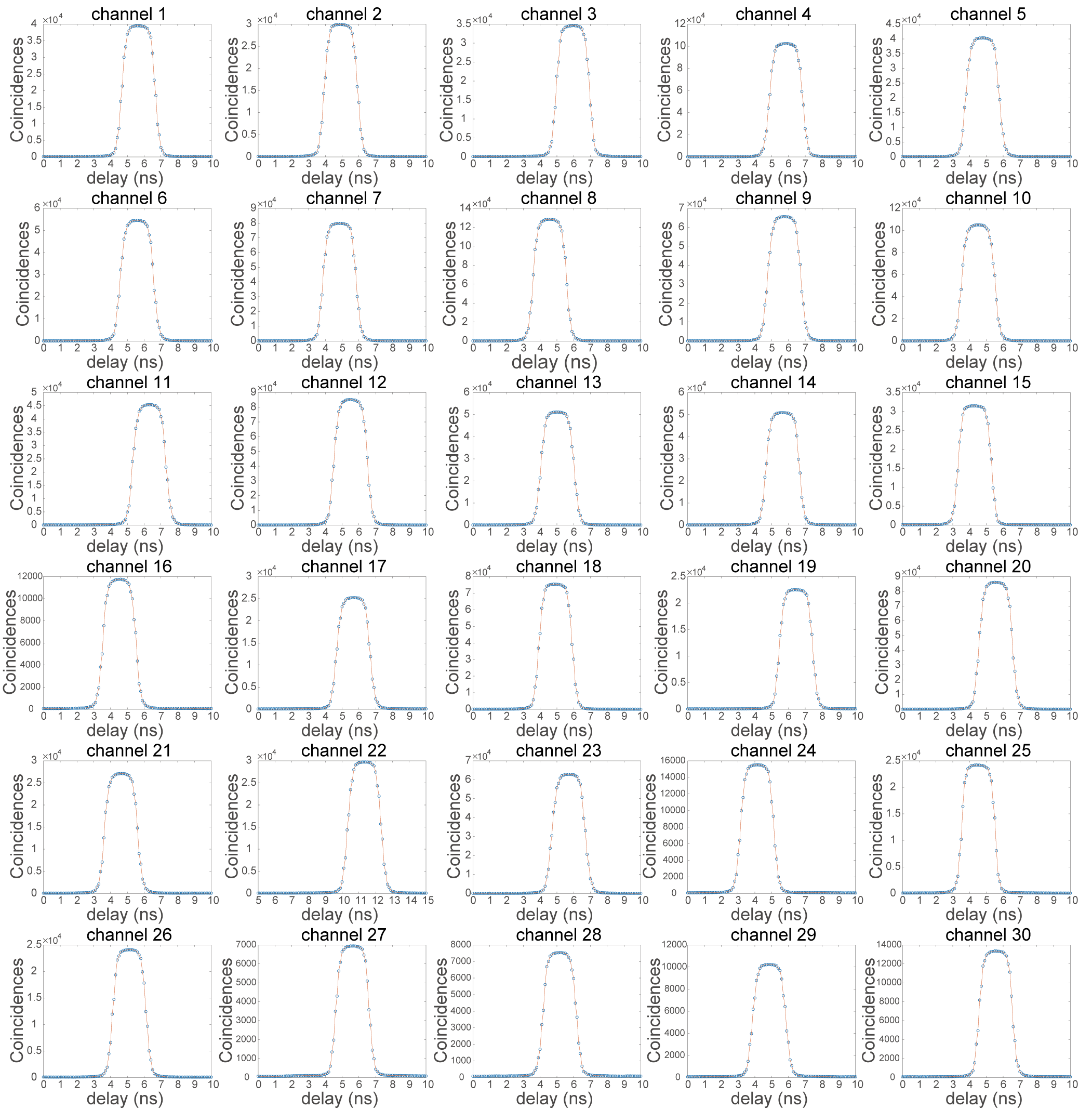}
		\caption{\textbf{Delay for all 30 channels with the time-of-flight technique.} Through using heralded single photons and the time-of-flight coincidence, we get the delay of all 30 channels accurately, and then determine the appropriate 4-fold coincidence time window.}
		\label{s1}
	\end{figure}

\begin{figure}[htbp]
		\centering
		\includegraphics[width=1.0\linewidth]{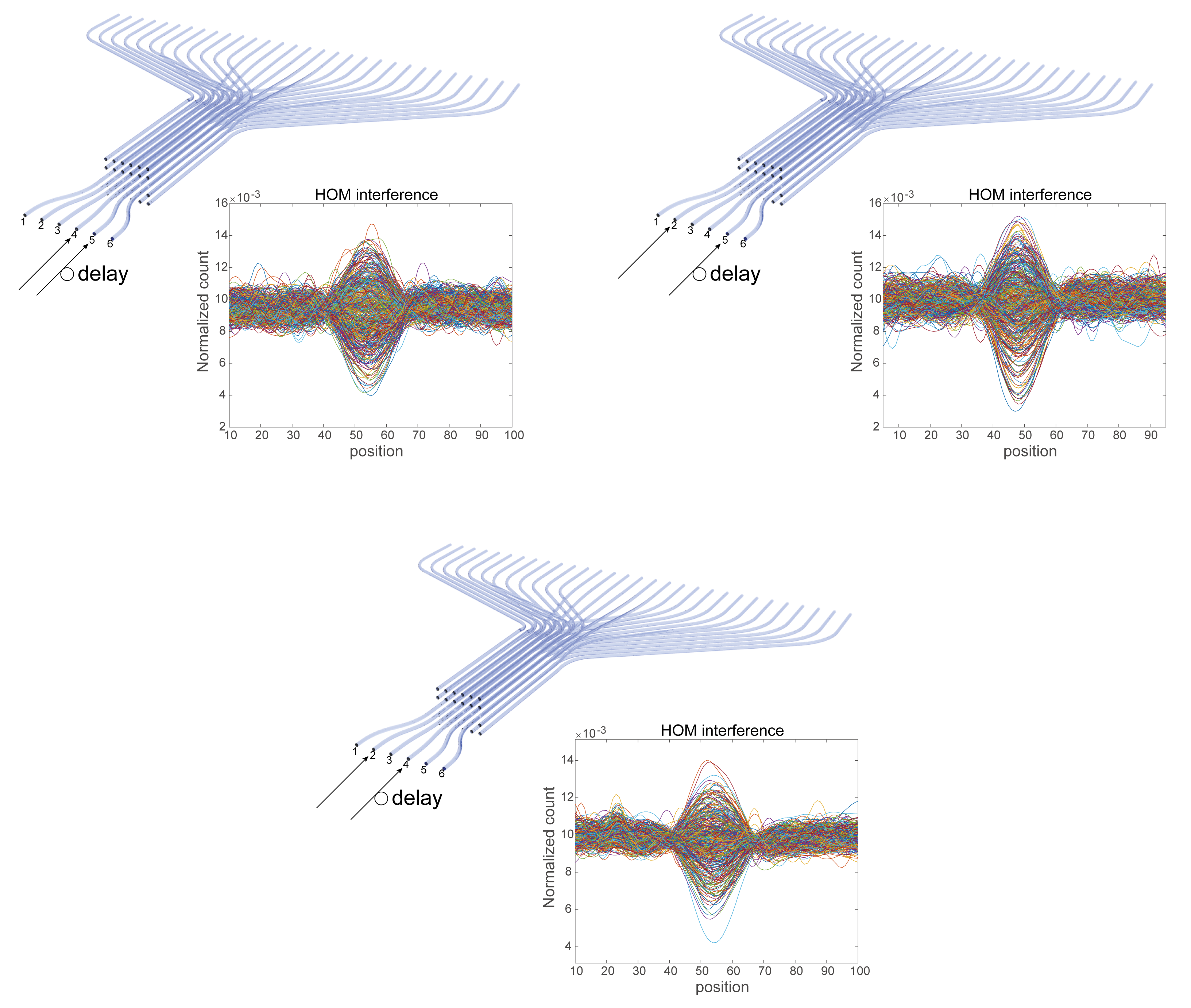}
		\caption{\textbf{All the HOM interferences datas during the scattering matrix characterization.} We measure the HOM interferences using the selected input ports 2,4,5 in order to obtain the phase relationship between different output modes. Each group contains 435 HOM interferences.}
		\label{s2}
	\end{figure}

\begin{figure}[htbp]
		\centering
		\includegraphics[width=1\linewidth]{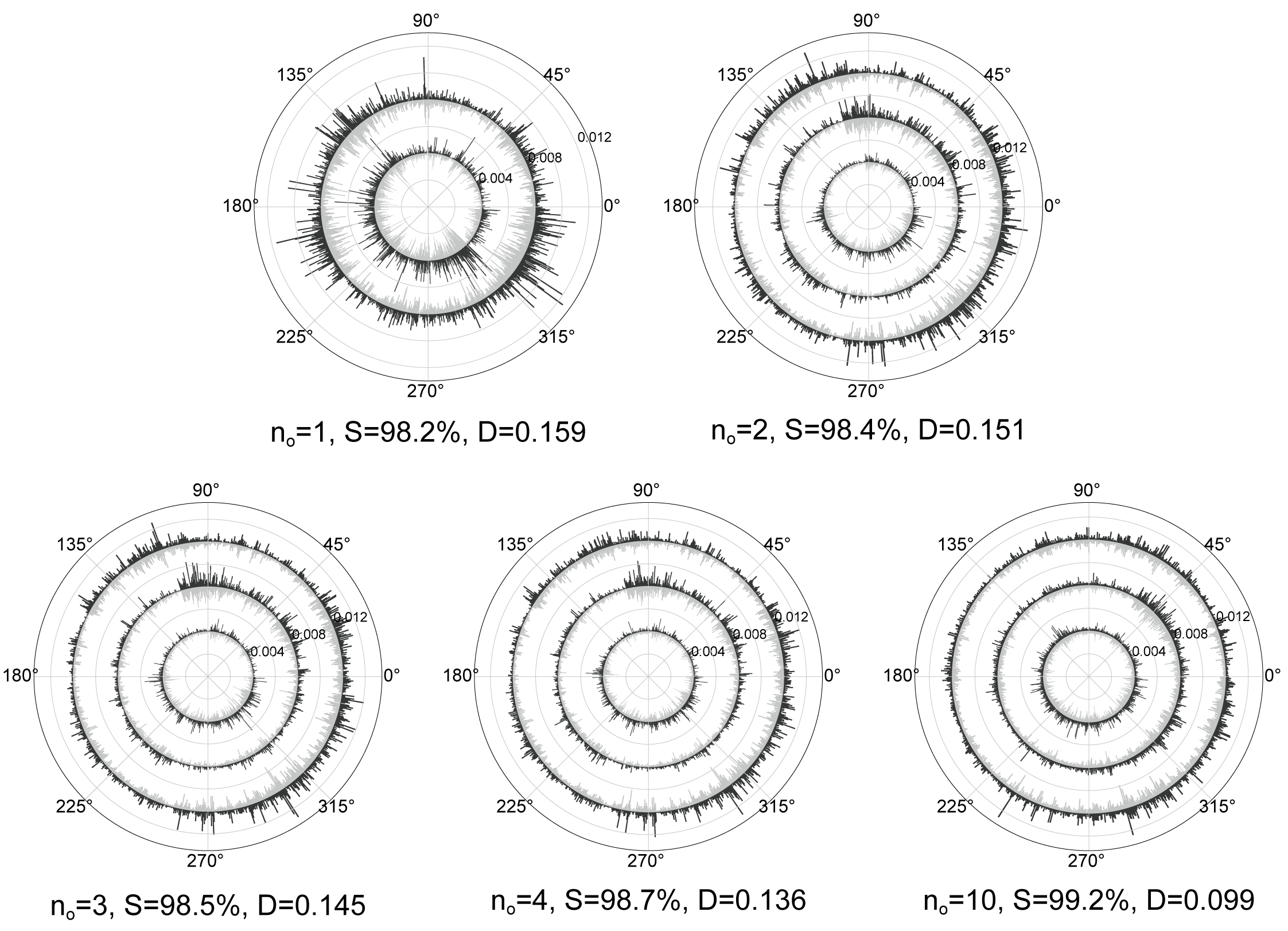}
		\caption{\textbf{Experimental comparison of the timestamp reconstruction and the counting reconstruction.}  The results of the timestamp reconstruction and the counting reconstruction under different $n_o$ which is the number of occurrences. We calculate and show all the similarity S which is defined as $S=\sum_{i}\sqrt{p_ic_i}$ and the total variation distance D which is defined as $D=(1/2)\sum_{i}\left|p_i-c_i\right|$. $p_i$ (dark grey) represents the timestamp reconstruction results and $c_i$ (light grey) represents the counting reconstruction results.}
		\label{s3}
	\end{figure}

\begin{figure}[htbp]
		\centering
		\includegraphics[width=1\linewidth]{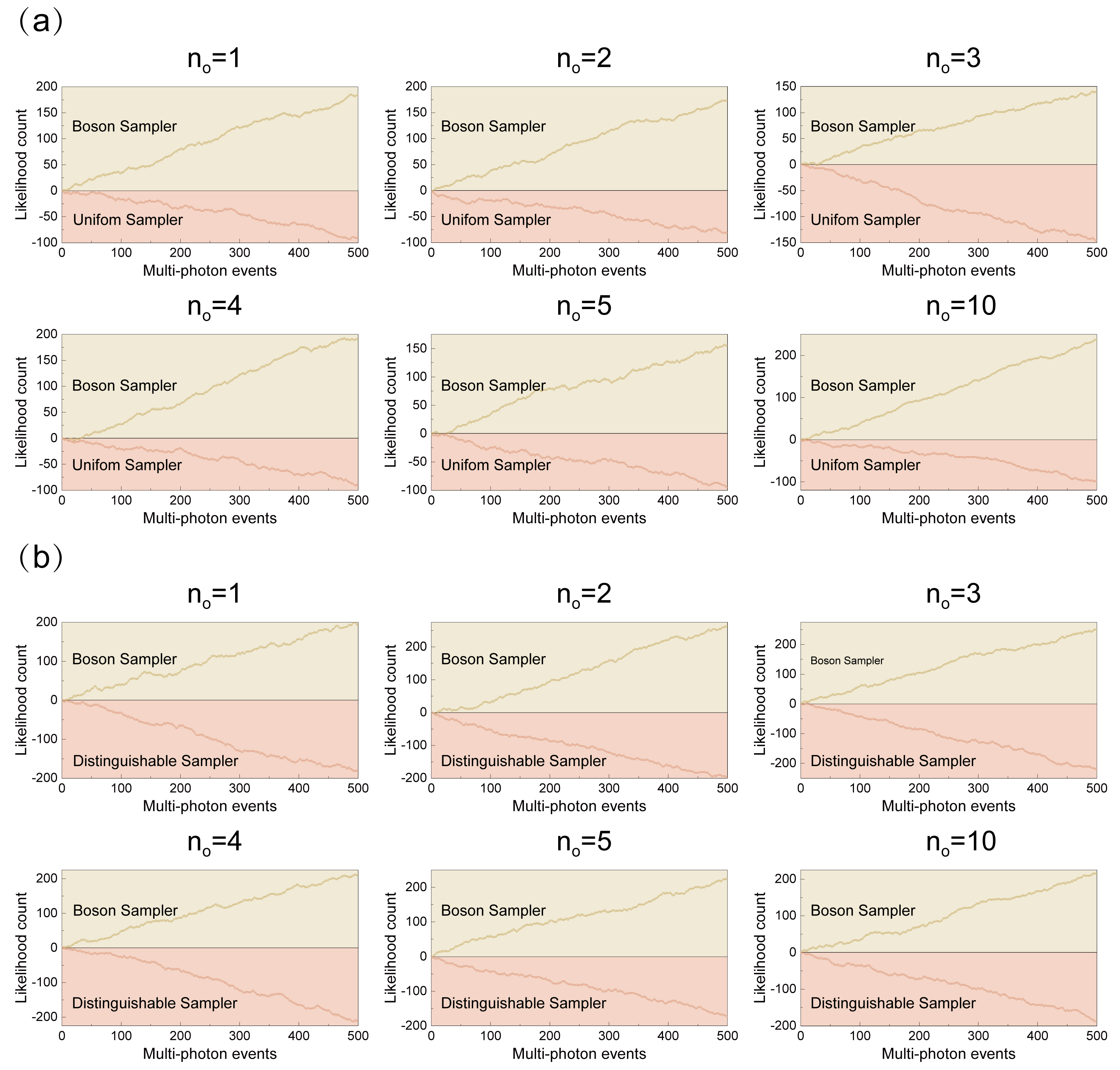}
		\caption{\textbf{Validation of the experimental boson sampler with the datas in Fig.4(c-h).} The experimental datas are verified both the row-norm estimator test and the likelihood ratio test. $n_o$ represent different occurrence times.}
		\label{s4}
	\end{figure}

	We show more preparation work and the experimental results in this section, including the delay for all 30 channels with the time-of-flight technique in Fig.\ref{s1}, all the HOM interferences datas in Fig.\ref{s2}, the experimental comparison of the timestamp reconstruction and the counting reconstruction with different $n_o$ (the number of occurrences) in Fig.\ref{s3}, and the validition which is the supplement to fig.4(c-h) in Fig.\ref{s4}.
	
	\clearpage

	\clearpage

	\clearpage

	\clearpage

\end{document}